\documentclass[sort&compress,a4paper,oneside,3p,12pt]{melsarticle}
\usepackage{verbatim}
\usepackage{amssymb}
\usepackage{amsmath,amssymb,amsfonts,graphicx,float,multicol,fleqn}
\usepackage{multirow}

\begin{document}

\begin{frontmatter}

\title{An approximate solution of the MHD Falkner-Skan flow by Hermite functions pseudospectral method}

\author[a]{K. Parand\corref{cor}}
\cortext[cor]{Corresponding author. Tel:+98 21 22431653; Fax:+98 21 22431650.}
\ead{k\_parand@sbu.ac.ir}
\fntext[a]{Member of research group of Scientific Computing.}
\author{A.R. Rezaei}
\ead{alireza.rz@gmail.com}
\author{S.M. Ghaderi}
\ead{ghaderi@khayam.ut.ac.ir}
\address{Department of Computer Sciences, Shahid Beheshti University, G.C., Tehran, Iran}

\begin{abstract}
Based on a new approximation method, namely pseudospectral method, a solution for the three order nonlinear ordinary differential laminar boundary layer Falkner-Skan equation has been obtained on the semi-infinite domain. The proposed approach is equipped by the orthogonal Hermite functions that have perfect properties to achieve this goal. This method solves the problem on the semi-infinite domain without truncating it to a finite domain and transforming domain of the problem to a finite domain. In addition, this method reduces solution of the problem to solution of a system of algebraic equations. We also present the comparison of this work with numerical results and show that the present method is applicable.
\end{abstract}


\end{frontmatter}

% \linenumbers

\section{Introduction}
\label{intro}
There are many problems in science and engineering arising in unbounded domains.
Spectral methods are famous ways to solve these kinds of problems.
The most common approach on spectral methods, that is used in this paper too, is through the use of polynomials/functions that are orthogonal over unbounded domains, such as the Hermite and the Laguerre polynomials/functions \cite{Coulaud,Funaro.Kavian,Funaro.Appl. Numer. Math.1990,Guo.Math. Comp.1999,Guo.num2000,Maday,Shen,Siyyam}.

Guo \cite{Guo.J. Math. Anal. Appl.1998,Guo.com2000,Guo.J. Math. Anal. Appl.2000} proposed a method that proceeds by mapping the original problem in an unbounded domain to a problem in a bounded domain, and then using Jacobi polynomials to approximate the resulting problems.

Another method is replacing infinite domain with $[-L, L]$ and semi-infinite interval with $[0, L]$ by choosing $L$, sufficiently large.
This method is called domain truncation \cite{BoydBook}.

Using rational approximation is another effective direct approach for solving such problems.
Christov \cite{Christov.SIAM J. Appl. Math.1982} and Boyd \cite{Boyd.J. Comput. Phys.1987(69),Boyd1987} developed some spectral methods on unbounded intervals by using mutually orthogonal systems of rational functions.
Boyd \cite{Boyd1987} defined a new spectral basis, named rational Chebyshev functions on the semi-infinite interval, by mapping to the Chebyshev polynomials.
Guo et al. \cite{Guo.sci2000} introduced a new set of rational Legendre functions which are mutually orthogonal in $L^2(0,+\infty)$.
They applied a spectral scheme using the rational Legendre functions for solving the Korteweg-de Vries equation on the half line.
Boyd et al. \cite{Boyd2003} applied pseudospectral methods on a semi-infinite interval and compared rational Chebyshev, Laguerre and mapped Fourier sine.

Parand et al. \cite{Parand.Appl. Math. Comput.2004,Parand.Int. J. Comput. Math.2004,Parand.Phys. Scripta2004,Parand.Phys.let.A,Parand.CAM,Parand.JCP} applied spectral method to solve nonlinear ordinary differential equations on semi-infinite intervals. Their approach was based on rational Tau and pseudospectral methods.

The concept of boundary layer flows \cite{Schlichting} of an incompressible fluid has several engineering applications, such as aerodynamic extrusion of plastic sheets and cooling of a metallic plate in a cooling bath, cooling and heating of the plate by free convection current \cite{Ogulu.Makinde}. Such flows are also encountered in the glass and polymer industries.
The effect of thermal radiation absorption on an unsteady free convective flow past a
vertical plate has been studied in the presence of a magnetic field and constant wall heat
flux by \cite{Ogulu.Makinde} also Makinde \cite{MakindeHFF2009} investigated numerically the hydromagnetic mixed convection flow of an incompressible viscous electrically conducting fluid and mass transfer over a vertical porous plate with constant heat flux embedded in a porous medium by Newton-Raphson shooting method along with fourth-order Runge-Kutta integration algorithm. The case of mixed-convective boundary layer flow past a vertical
porous plate embedded in a saturated porous medium with a constant heat flux and mass transfer in the presence of a magnetic field and with heat absorption has been studied by \cite{Makinde.Sibanda2008}.

The steady laminar flow passing a fixed wedge was first analyzed in the early 1930s by Falkner and Skan \cite{Falkner-Skan} to illustrate the application of Prandtl's boundary layer theory.
The Magnetohydrodynamic (MHD) flows as boundary layer flows have a great significance both from a mathematical as well as a physical standpoint. Such flows are very important in electromagnetic propulsion.
The MHD systems are used effectively in many applications including power generators, pumps, accelerators, electrostatic filters, droplet filters, the design of heat exchangers, the cooling of reactors etc. \cite{Sutton-Sherman}.
\subsection{MHD flows}
Consider the boundary layer flow of an electrically conducting viscous fluid. A magnetic field $B(x)$ acts transversely to the flow. The induced magnetic field is neglected by choosing small magnetic Reynolds number assumption. Furthermore, the electric field is absent. The relevant problem is
\begin{eqnarray}
\frac{\partial u}{\partial x}+\frac{\partial v}{\partial y}=0,
\end{eqnarray}
\begin{eqnarray}\label{Eq.partial-falkner}
u\frac{\partial u}{\partial x}+v\frac{\partial u}{\partial y}=U\frac{\mathrm{d} u}{\mathrm{d} x}+\nu\frac{\partial^2 v}{\partial y^2}-\frac{\sigma B^2}{\rho}(u-U),
\end{eqnarray}
\begin{equation}\label{Eq.partial-falkner-con}
u=0,~~~~v=0 ~ at ~ y=0,\\
u=U(x), ~~~ as ~ y\rightarrow \infty,
\end{equation}
where \cite{Schlichting}
\begin{eqnarray}
U(x)=ax^m
\end{eqnarray}
and \cite{chiam}
\begin{eqnarray}
B(x)=B_{0}x^{(m-1)/2},
\end{eqnarray}
in which $u$ and $v$ are the velocity components, $U$ is the inherent
characteristic velocity, $\nu$ is a kinematic viscosity, $\sigma$ is the electrical
conductivity, $\rho$ is the fluid density, $B$ and $B_0$ are the magnetic field and externally imposed magnetic field in the $y$-direction respectively.

Defining
\begin{eqnarray}
\tau=\sqrt{\frac{m+1}{2}}\sqrt{\frac{U}{\nu x}}y,~~~~\psi=\sqrt{\frac{2}{m+1}}\sqrt{\nu x U}f(\tau),
\end{eqnarray}
\begin{eqnarray}
u=Uf'(\tau),~~~~~v=-\sqrt{\frac{m+1}{2}}\sqrt{\frac{\nu U}{x}}[f+\frac{m-1}{m+1}\tau f'],
\end{eqnarray}
the continuity equation is identically satisfied and Eq. \ref{Eq.partial-falkner} and
boundary conditions \ref{Eq.partial-falkner-con} reduce to the following form
\begin{eqnarray}\label{Falkner.main}
\frac{\mathrm{d}^{3}f}{\mathrm{d}\tau^3}+f\frac{\mathrm{d}^{2}f}{\mathrm{d}\tau^2}+\beta[1-(\frac{\mathrm{d}f}{\mathrm{d}\tau})^2]-M^2(\frac{\mathrm{d}f}{\mathrm{d}\tau}-1)=0,
\end{eqnarray}
\begin{eqnarray}\label{Falkner.Bound}
f(0)=f'(0)=0, \quad f'(+\infty)=1,
\end{eqnarray}
where $\beta=\frac{2m}{m+1}$ and $M^2=2\sigma B_{0}^2/\rho a(1+m)$. Now our interest is to
find solution of Eq. (\ref{Falkner.main}) for wedge in the accelerated flow $(m>0,\beta>0)$ and decelerated flow $(m<0,\beta<0)$ with separation.

MHD boundary layer flows have been studied by several researchers. Yih \cite{Yih} and Ishak et al. \cite{ishak-nazar-pop} transformed the partial differential boundary layer equations into the nonsimilar boundary layer equations and a system of ordinary differential equations respectively, then they used Keller box method to solve them.
Hayat et al. \cite{hayat-hussain-javad}, Rashidi \cite{Rashidi} and Abbasbandy et al. \cite{Addasbandi-Hayat1,Addasbandi-Hayat2} solved MHD boundary layer flow by modified Adomian decomposition method, DTM-Pad\'{e}, Homotopy analysis method and Hankel-Pad\'{e} respectively.

The purpose of this paper is to develop a pseudospectral method that is more direct and simpler than the other spectral methods. The method of this paper differs from previous methods that have been used because it approximates the solution directly. Thus, it does not truncate the semi-infinite domain to a finite domain, does not introduce the unknown finite boundary, does not impose an asymptotic condition at this unknown boundary, and does not use transformation to map the physical domain.

This paper is organized as follows:\\
In section \ref{HFCmethod}, the properties of Hermite functions and the way to construct the pseudospectral approach for this type equation are described. In section \ref{SolvingFSF} the proposed method is applied to solve MHD Falkner-Skan equation, and a comparison is made with existing numerical solutions that were reported in other researches, and we have a discussion in the final section.
\section{\bf{Hermite functions pseudospectral method}}\label{HFCmethod}
Spectral methods have been successfully applied in the approximation of differential
boundary value problems defined in unbounded domains. For problems whose solutions are
sufficiently smooth, they exhibit exponential rates of convergence/spectral accuracy. We can apply different spectral methods that are used to solve problems in semi-infinite domains. One of these approaches is using Laguerre and Hermite polynomials/functions \cite{Funaro.Kavian,Guo.Math. Comp.1999,Guo.num2000,Maday,Shen,Siyyam,GuoShenXu2003,Bao.Shen,Guo.Xu}. Guo \cite{Guo.num2000} suggested a Laguerre-Galerkin method for the Burgers equation and Benjamin-Bona-Mahony (BBM) equation on a semi-infinite interval.
In \cite{Shen} Shen proposed spectral methods using Laguerre functions and analyzed for model elliptic equations on regular unbounded domains.
Siyyam \cite{Siyyam} applied two numerical methods for solving initial value problem
differential equations using the Laguerre Tau method.
Maday et al. \cite{Maday} proposed a Laguerre type spectral method for solving partial
differential equations.
Funaro and Kavian \cite{Funaro.Kavian} considered some algorithms by using Hermite functions. Recently Guo \cite{Guo.Math. Comp.1999} developed the spectral method by using Hermite polynomials. However, it is not easy to perform the quadratures in unbounded domains, which are used in the Hermite spectral approximations. So the Hermite pseudospectral method is more preferable in actual calculations. Guo \cite{Guo.Xu} developed the Hermite pseudospectral method for the Burgers equation on the whole line. Guo et al. \cite{GuoShenXu2003} considered spectral and pseudospectral approximations using Hermite functions for PDEs on the whole line to approximate the Dirac equation. Bao and Shen \cite{Bao.Shen} proposed a generalized-Laguerre-Hermite pseudospectral method for computing symmetric and central vortex states in Bose-Einstein condensates (BECs) in three dimensions with cylindrical symmetry.

Pseudospectral methods has become increasingly popular for solving differential
equations and also it is very useful in providing highly accurate solutions to differential equations.
In this paper, we employ the Hermite functions pseudospectral method which we denote HFP to solve MHD Falkner-Skan initial value problem directly.
%%%%%%%%%%%%%%%%%%%%%%%%%%%%%%%%%%%%%
\subsection{Properties of Hermite functions}
In this section, we detail the properties of the Hermite functions that will be used to construct the HFP method.
First we note that the Hermite polynomials are generally not suitable in practice due to their wild asymptotic behavior at infinities \cite{ShenWang2008,SzegoBook}.\\
Hermite polynomials $H_n(x)$, $n\geq0$, are the eigenfunctions of the singular
Sturm-Liouville problem in
\begin{equation}\nonumber
H^{''}_n(x)-2xH^{'}_n(x)+2nH_n(x)=0
\end{equation}
Hermite polynomials with large $n$ can be written in direct formula as follow:
%--------------------------------------------------------------------------------
\begin{align}\nonumber
H_{n}(x)&\sim\frac{\Gamma(n+1)}{\Gamma(n/2+1)}e^{x^2/2}\cos{(\sqrt{2n+1}x-\frac{n\pi}{2})}\\\nonumber
&\sim n^{n/2}e^{x^2/2}\cos(\sqrt{2n+1}x-\frac{n\pi}{2}).
\end{align}
Hence, we shall consider the so called Hermite functions.
The normalized Hermite functions of degree $n$ are defined by
\begin{eqnarray}\nonumber
\widetilde{H}_{n}(x)=\frac{1}{\sqrt{2^n n!}}e^{-x^2/2}H_{n}(x), \quad n\geq 0, x\in \mathbb{R}.
\end{eqnarray}
Clearly, \{$\widetilde{H}_{n}$\} is an orthogonal system in $L^2(\mathbb{R})$, i.e.,
\begin{eqnarray}\nonumber
\int^{+\infty}_{-\infty}\widetilde{H}_{n}(x)\widetilde{H}_{m}(x)dx=\sqrt{\pi}\delta_{mn}.
\end{eqnarray}
where $\delta_{nm}$ is the Kronecker delta function.\\
In contrast to the Hermite polynomials, the Hermite functions are well behaved with the decay property:
\begin{eqnarray}\nonumber
|\widetilde{H}_{n}(x)|\longrightarrow 0, \quad as \quad |x|\longrightarrow \infty,
\end{eqnarray}
and the asymptotic formula with large $n$ is
\begin{eqnarray}\nonumber
\widetilde{H}_{n}(x)\sim n^{-\frac{1}{4}}\cos(\sqrt{2n+1}x-\frac{n\pi}{2})
\end{eqnarray}
The three-term recurrence relation of Hermite polynomials implies
\begin{eqnarray}\nonumber
&\widetilde{H}_{n+1}(x)=x\sqrt{\frac{2}{n+1}}\widetilde{H}_{n}(x)-\sqrt{\frac{n}{n+1}}\widetilde{H}_{n-1}(x),\quad n\geq{1},\\\nonumber
&\noindent\widetilde{H}_{0}(x)=e^{-x^2/2}, \quad \widetilde{H}_{1}(x)=\sqrt{2}xe^{-x^2/2}.
\end{eqnarray}
Using recurrence relation of Hermite polynomials and the above formula leads to
\begin{align}\nonumber
\widetilde{H}'_{n}(x)&=\sqrt{2n}\widetilde{H}_{n-1}(x)-x\widetilde{H}_{n}(x)\\\nonumber
&=\sqrt{\frac{n}{2}}\widetilde{H}_{n-1}(x)-\sqrt{\frac{n+1}{2}}\widetilde{H}_{n+1}(x).
\end{align}
and this implies
\begin{eqnarray}\nonumber
\int_{\mathbb{R}}{\widetilde{H}'_n(x)\widetilde{H}'_m(x)dx}=
\begin{cases}
-\frac{\sqrt{n(n-1)\pi}}{2}, & m=n-2, \\
\sqrt{\pi}(n+\frac{1}{2}), & m=n,\\
-\frac{\sqrt{(n+1)(n+2)\pi}}{2}, & m=n+2, \\
0, & \text{otherwise}.
\end{cases}
\end{eqnarray}
Let us define
\begin{eqnarray}\nonumber
\widetilde{P}_N:=\{u:u=e^{-x^2/2}v,\forall v\in{P_N}\}.
\end{eqnarray}
where $P_N$ is the set of all Hermite polynomials of degree at most $N$.\\
We now introduce the Gauss quadrature associated with the Hermite functions approach.\\
Let $\{x_j\}_{j=0}^{N}$ be the Hermite-Gauss nodes and define the weights
\begin{eqnarray}\nonumber
\widetilde{w}_j=\frac{\sqrt{\pi}}{(N+1)\widetilde{H}_{N}^{2}(x_j)}, \quad 0\leq j\leq N.
\end{eqnarray}
Then we have
\begin{eqnarray}\nonumber
\int_{\mathbb{R}}p(x)dx=\sum_{j=0}^{N}p(x_j)\widetilde{w}_j, \quad \forall p\in \widetilde{P}_{2N+1}.
\end{eqnarray}
For a more detailed discussion of these early developments, see the \cite{ShenTangHighOrder,ShenTangWang}.

\subsection{Approximations by Hermite functions}
Let us define $\Lambda:=\{x|-\infty<x<\infty\}$
and
\begin{eqnarray}\nonumber
\mathcal{H}_N=span\{\widetilde{H}_0(x),\widetilde{H}_1(x),...,\widetilde{H}_ N(x)\}
\end{eqnarray}
The $L^{2}{(\Lambda)}$-orthogonal projection $\tilde{\xi}_N:L^{2}{(\Lambda)}\longrightarrow \mathcal{H}_N$ is a mapping in a way that for any $v\in L^{2}{(\Lambda)}$,
\begin{eqnarray}\nonumber
<\tilde{\xi}_N{v}-v,\phi>=0, \quad \forall \phi\in \mathcal{H}_N
\end{eqnarray}
or equivalently,
\begin{eqnarray}\nonumber
\tilde{\xi}_N{v(x)}=\sum_{l=0}^{N}\tilde{v}_{l}\widetilde{H}_l(x).
\end{eqnarray}
To obtain the convergence rate of Hermite functions we define the space $H_{A}^{r}(\Lambda)$ defined by
\begin{eqnarray}\nonumber
H^{r}_{A}{(\Lambda)}=\{v|v \text{ is measurable on } \Lambda \text{ and }{\|v\|}_{r,A}<\infty\},\nonumber
\end{eqnarray}
and equipped with the norm $\|v\|_{r,A}=\|A^{r}v\|$. For any $r > 0$, the space $H^{r}_{A}{(\Lambda)}$ and its norm are defined by space interpolation. By induction, for any non-negative integer $r$,
\begin{eqnarray}\nonumber
A^{r}v(x)=\sum_{k=0}^{r}(x^2+1)^{(r-k)/2}p_k(x){\partial}_{x}^{k}v(x),
\end{eqnarray}
where $p_k(x)$ are certain rational functions which are bounded uniformly on $\Lambda$. Thus,
\begin{eqnarray}\nonumber
\|v\|_{r,A}\leq c \left(\sum_{k=0}^{r}\parallel(x^2+1)^{(r-k)/2}p_k(x)\partial_{x}^{k}v\parallel \right)^{1/2}.
\end{eqnarray}
%\begin{theorem}
{\bf {Theorem :}}
For any $v \in H^{r}_{A}(\Lambda)$, $r\geq1$ and $0\leq \mu \leq r$,
\begin{equation}
\parallel{\tilde{\xi}}_{N}v-v{\parallel_{\mu}\leq cN^{1/3+(\mu-1)/2}\parallel v\parallel}_{r,A}.
\end{equation}
%\end{theorem}
{\it Proof}. A complete proof is given by Guo et al. \cite{GuoShenXu2003}. also same theorem has been proved by Shen et al. \cite{ShenWang2008}.
\subsection{Hermite functions transform}
As mentioned before, Falkner-Skan equation is defined on the interval $(0,+\infty)$; but we know properties of Hermite functions are derived in the infinite domain $(-\infty,+\infty)$.\\
Also we know approximations can be constructed for infinite, semi-infinite and finite intervals.
One of the approaches to construct approximations on the interval $(0,+\infty)$ which is used in this paper, is to use a mapping, that is a change of variable of the form
\begin{equation}\nonumber
\omega=\phi(z)=\frac{1}{k}\ln(z).
\end{equation}
where $k$ is a constant.\\
The basis functions on $(0,+\infty)$ are taken to be the transformed Hermite functions,
\begin{eqnarray}\nonumber
\widehat{H}_n(x)\equiv \widetilde{H}_n(x)\circ \phi(x)= \widetilde{H}_n(\phi(x)).
\end{eqnarray}
where $\widetilde{H}_n(x)\circ \phi(x)$ is defined by $\widetilde{H}_n(\phi(x))$. The inverse map of $\omega=\phi(z)$ is
\begin{eqnarray}\label{inverseTransform}
z=\phi^{-1}(\omega)=e^{k\omega}.
\end{eqnarray}
Thus we may define the inverse images of the spaced nodes ${ {\{{x_j}}\}_{-\infty}^{+\infty} }$ as
\begin{eqnarray}\nonumber
\Gamma=\{\phi^{-1}(t): -\infty< t <+\infty\}=(0,+\infty)
\end{eqnarray}
and
\begin{eqnarray}\nonumber
\tilde{x}_j=\phi^{-1}(x_j)=e^{k{x_j}},\quad j=0,1,2,...
\end{eqnarray}
Let $w(x)$ denote a non-negative, integrable, real-valued function over the interval $\Gamma$.
We define
\begin{eqnarray}\nonumber
L^2_w(\Gamma)=\{v:\Gamma\rightarrow \mathbb{R}\mid v \textrm{ is measurable and}\parallel v{\parallel}_w<\infty \}
\end{eqnarray}
where
\begin{eqnarray}\nonumber
\parallel v{\parallel}_w=\left(\int_{0}^\infty\mid v(x)\mid ^2w(x)\mathrm{d}x\right)^{\frac{1}{2}},
\end{eqnarray}
is the norm induced by the inner product of the space $L^2_w(\Gamma)$,
\begin{equation}\label{Eq.inner product definition}
<u,v>_w=\int_{0}^{\infty}u(x)v(x)w(x)\mathrm{d}x.
\end{equation}
Thus $\{\widehat{H}_n(x)\}_{n\in \mathbb{N}}$ denotes a system which is mutually orthogonal under (\ref{Eq.inner product definition}), i.e.,
\begin{eqnarray}\nonumber
< \widehat{H}_n(x),\widehat{H}_m(x)>_{w(x)}=\sqrt{\pi}\delta_{nm},
\end{eqnarray}
where $w(x)=1/x$ and $\delta_{nm}$ is the Kronecker delta function. This system is complete in $L^2_w(\Gamma)$. For any function $f\in L^2_w(\Gamma)$ the following expansion holds

\begin{eqnarray}\nonumber
f(x)\cong \sum_{k=0}^{N}f_k \widehat{H}_k(x),
\end{eqnarray}
%\begin{eqnarray}\nonumber
%{\int_{0}^\infty} f(x)\cong \sum_{k=-N}^{+N}\frac{f_k}{\phi'(x)},
%\end{eqnarray}
with
\begin{eqnarray}\nonumber
f_k=\frac{<f(x),\widehat{H}_k(x)>_{w(x)}}{\parallel \widehat{H}_k(x){\parallel}_{w(x)}^2}.
\end{eqnarray}
Now we can define an orthogonal projection based on transformed Hermite functions as below:\\
Let
\begin{eqnarray}\nonumber
\widehat{\mathcal{H}}_N=span\{\widehat{H}_0(x),\widehat{H}_1(x),...,\widehat{H}_n(x)\}
\end{eqnarray}
The $L^{2}{(\Gamma)}$-orthogonal projection $\hat{\xi}_N:L^{2}{(\Gamma)}\longrightarrow\widehat{\mathcal{H}}_N$ is a mapping in a way that for any $y\in L^{2}{(\Gamma)}$,
\begin{eqnarray}\nonumber
<\hat{\xi}_N{y}-y,\phi>=0, \quad \forall \phi\in \widehat{\mathcal{H}}_N
\end{eqnarray}
or equivalently,
\begin{eqnarray}\label{operatorHFT}
\hat{\xi}_N{y(x)}=\sum_{i=0}^{N}\hat{a}_{i}\widehat{H}_i(x).
\end{eqnarray}

%\bigskip
\subsection{Domain scaling}
It has already been mentioned in \cite{Liu.Liu.Tang} that when using a spectral approach on the whole real line $\mathbb{R}$ one can possibly increase the accuracy of the computation by a suitable scaling of the underlying time variable $t$. For example, if $y$ denotes a solution of the ordinary differential equation, then the rescaled function is
$\tilde{y}(t)=y(\frac{t}{l})$, where $l$ is constant.\\
Domain scaling is used in several of the applications presented in the next section. For more detail we refer the reader to \cite{Tang.1993}.
\section{Solving Falkner-Skan wedge flow equation with Hermite functions}\label{SolvingFSF}
Here, we solve Falkner-Skan wedge flow equation for $m=-3/5$ with $M=5,10,15,20$ and $m=2$ with $M=5,10,50,100$ by pseudospectral method.

We note that the Hermite functions are not differentiable at the point $x=0$, therefore to approximate the solution of Eq. (\ref{Falkner.main}) with the boundary conditions Eq. (\ref{Falkner.Bound}) we multiply operator Eq. (\ref{operatorHFT}) by ${x^2}/(x+1)$ to satisfy $f(0)=f'(0)=0$ in Eq. (\ref{Falkner.Bound}) and also we construct a function $p(x)$ that satisfy $f'(+\infty)=1$ in Eq. (\ref{Falkner.Bound}).
This function is given by
\begin{equation}\label{px}
p(x)=\frac{x^2}{x+\lambda},
\end{equation}
where $\lambda$ is constant to be determined.\\
Therefore, the approximate solution of $f(x)$, in Eq. (\ref{Falkner.main}) with boundary conditions Eq. (\ref{Falkner.Bound}) is represented by
\begin{eqnarray}\nonumber
\widehat{\xi}_{N}f(x)= p(x)+\frac{x^2}{x+1}\hat{\xi}_Nf(x),
\end{eqnarray}
To apply the pseudospectral scheme, we construct the residual function by substituting $f(x)$ by $\widehat{\xi}_{N}f(x)$ in the Falkner-Skan Eq. (\ref{Falkner.main}):
\begin{align}
Res_l(x)&=\frac{\mathrm{d}^3\widehat{\xi}_Nf(x/l)}{\mathrm{d}x^3}+\left(\widehat{\xi}_{N}f(x/l)\right)\left(\frac{\mathrm{d}^2\widehat{\xi}_Nf(x/l)}{\mathrm{d}x^2}\right)\\ &+\beta\left(1-(\frac{\mathrm{d}\widehat{\xi}_Nf(x/l)}{\mathrm{d}x})^2\right)-M^2\left(\frac{\mathrm{d}\widehat{\xi}_Nf(x/l)}{\mathrm{d}x}-1\right),
\end{align}
where $l$ is a constant that is defined in domain scaling description, before.
$\widehat{\xi}_Nf(x)$ will be a good approximation of function $f(x)$, if the residual function is zero on the whole domain.
In other words, we should select coefficients $a_i$s so that the residual function tends to zero on most of the domain.
The pseudospectral scheme for Falkner-Skan equation is to find $\widehat{\xi}_Nf(x)$ such that
\begin{equation}\label{Eq.Falkner-Skan equation by pseudospetral}
Res_{l}(x_j)=0,\qquad j=0,\ldots,N+1,
\end{equation}
where the $x_j$s are $N+2$ transformed Hermite-Gauss nodes by Eq. \ref{inverseTransform}.
This generates a set of $N+2$ nonlinear equations that can be solved by Newton method for unknown coefficients $a_i$s and $\lambda$.

The physical quantities of interest which is represented by value of $f''(0)$ is the skin friction coefficient. The approximations of the $f''(0)$ obtained by this method and numerical value \cite{Asaithambi} for $m=-3/5$ with $M=5,10,15,20$ and $m=2$ with $M=5,10,50,100$ are listed in Tables \ref{Tab.Falkner-m3-5} and \ref{Tab.Falkner-m2} respectively.
%Compared to the results with exact values, our solution is more accurate.

\begin{figure}
\includegraphics[scale=0.3]{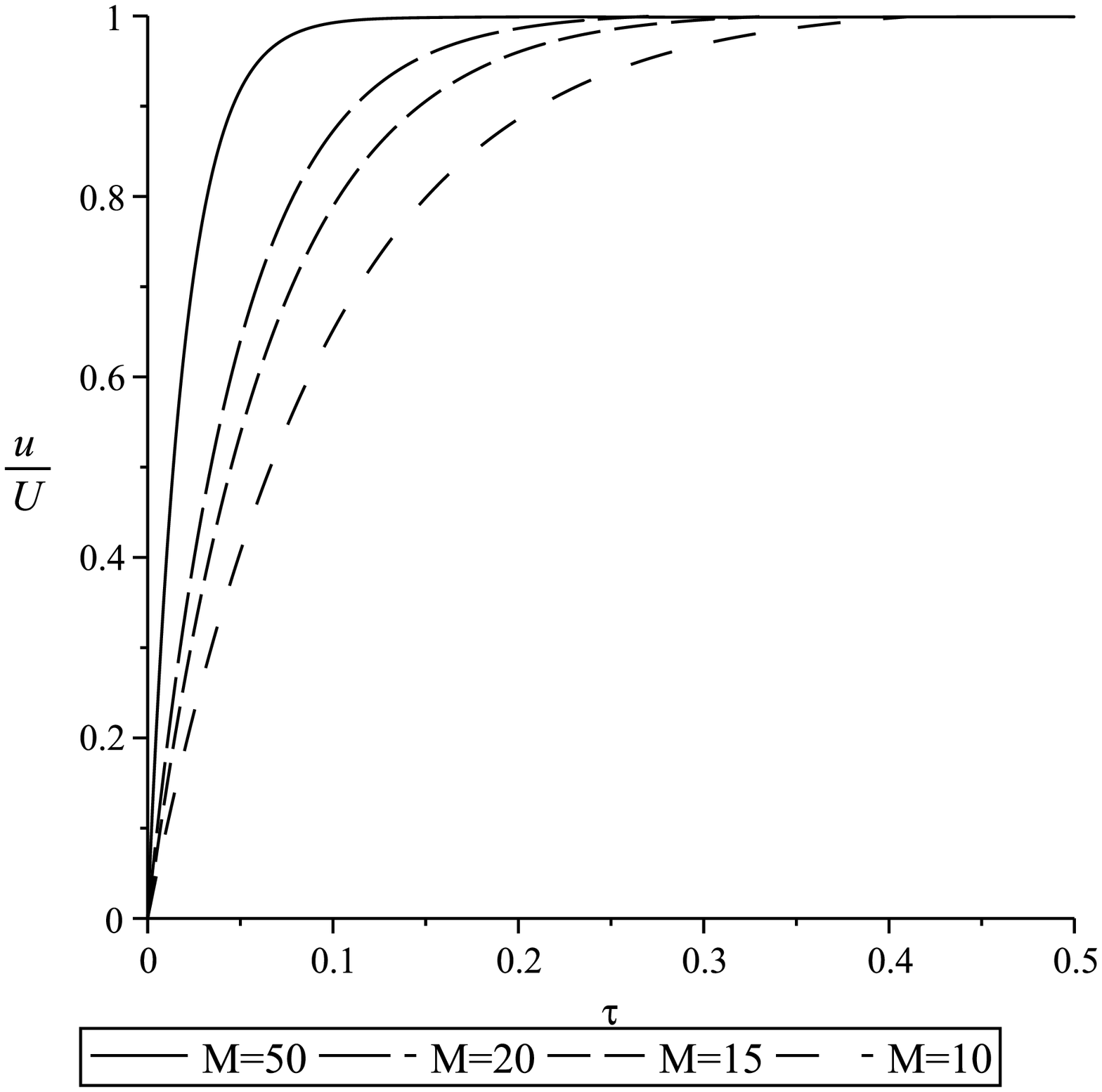}
\caption{The velocity boundary layer of the wedge with $m=-3/5$ and various $M$}
\label{figm3-5}
\end{figure}

\begin{figure}
\includegraphics[scale=0.3]{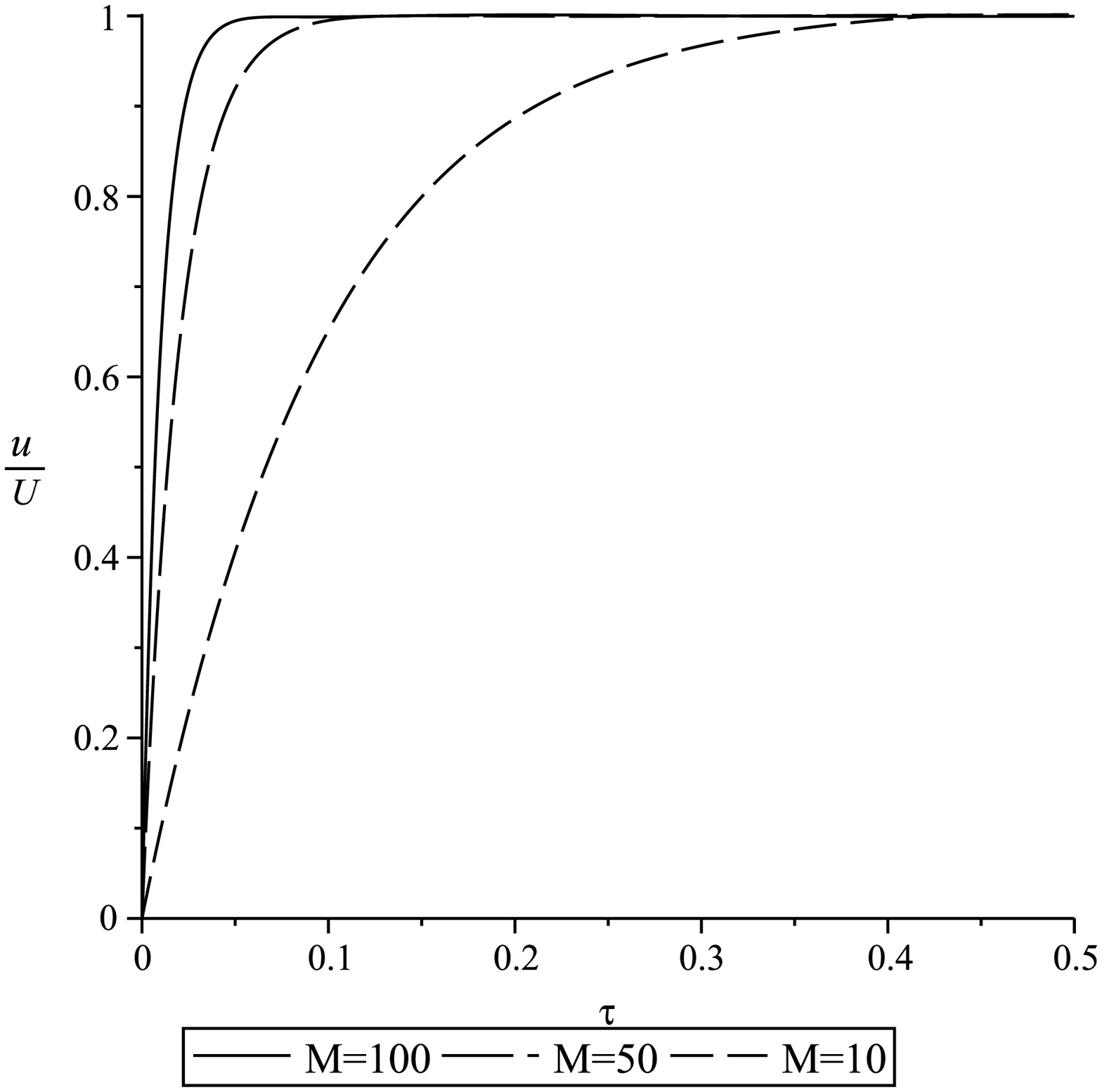}
\caption{The velocity boundary layer of the wedge with $m=2$ and various $M$}
\label{figm2}
\end{figure}

The velocity boundary layer of the wedge with $m=-3/5$ and $m=2$ for various various $M$ are shown in figures \ref{figm3-5} and \ref{figm2} respectively. The value of $f''(0)$ can be seen to increase with $M$. Hence, the presence of a magnetic field also increases the skin friction.

\begin{figure}
\includegraphics[scale=0.3]{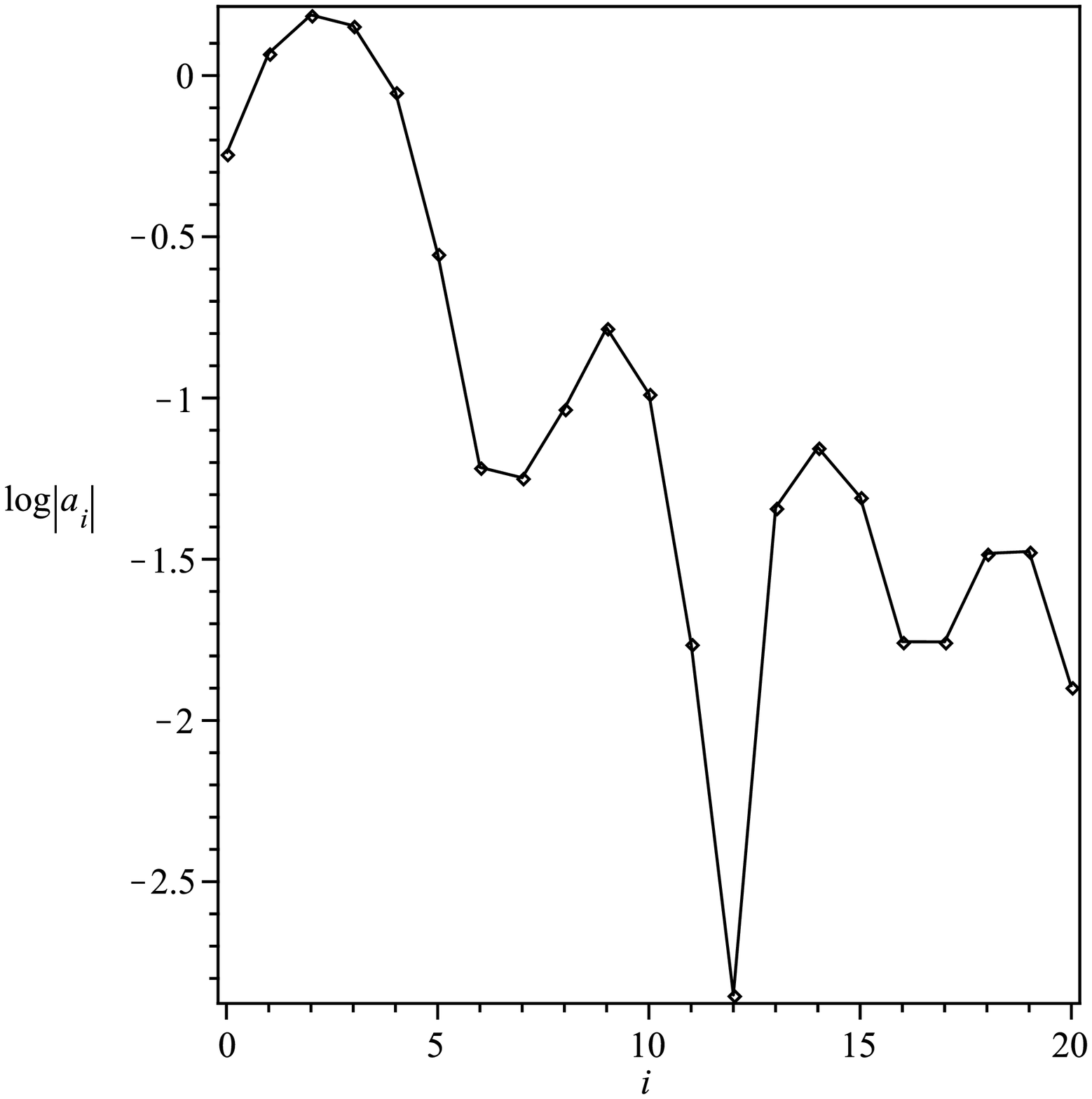}
\caption{Logarithmic graph of absolute coefficients $|a_i|$ of Hermite functions in the approximate solution for $m=-3/5$ and $M=50$}
\label{FigCoeffm3-5M50}
\end{figure}

\begin{figure}
\includegraphics[scale=0.3]{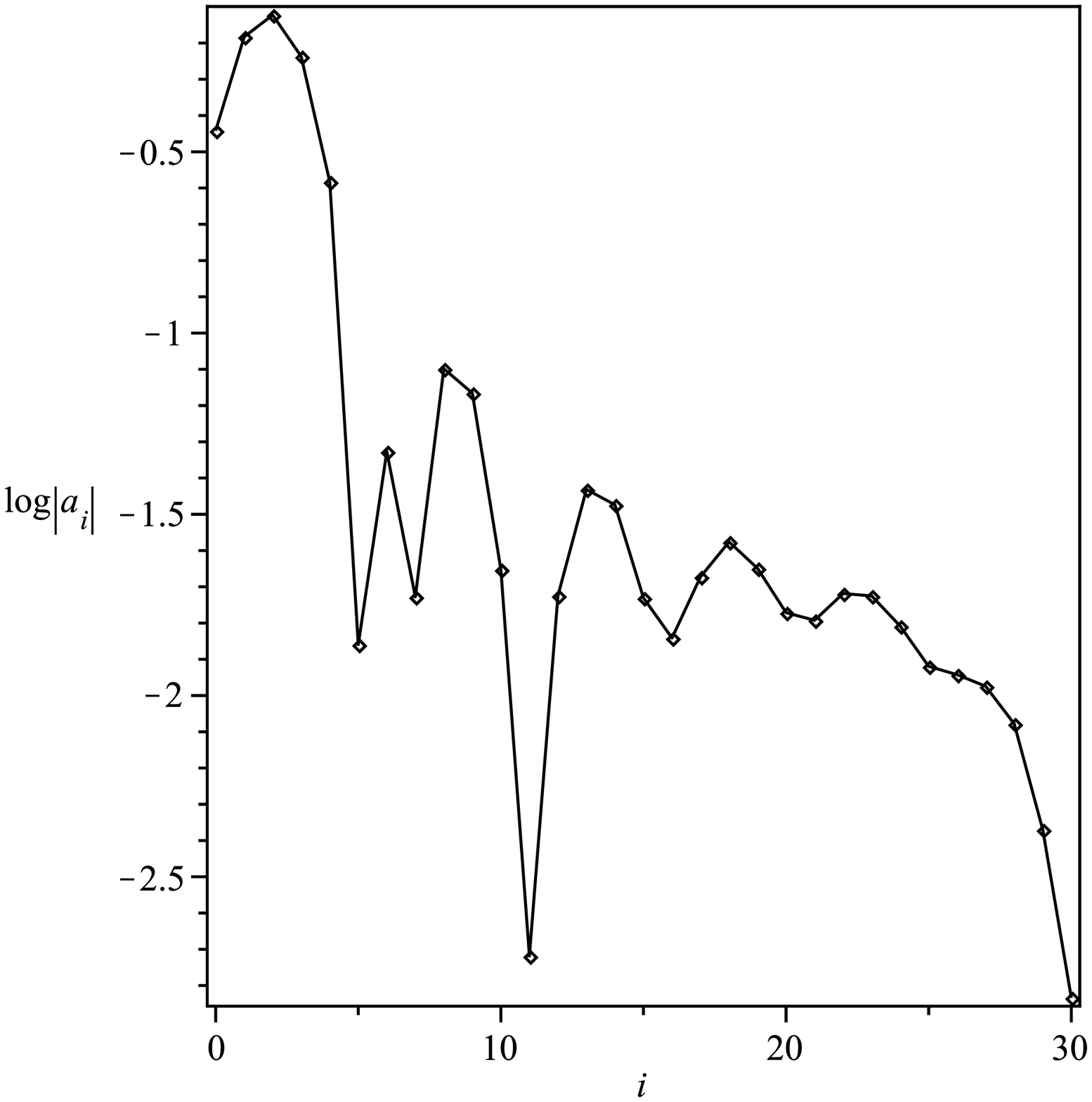}
\caption{Logarithmic graph of absolute coefficients $|a_i|$ of Hermite functions in the approximate solution for $m=2$ and $M=50$}
\label{FigCoeffm2M50}
\end{figure}

Logarithmic graphs of absolute coefficients $|a_i|$ of Hermite function in the approximate solutions for $m=-3/5$ with $M=50$ and $m=2$ with same $M$ are shown in Figures \ref{FigCoeffm3-5M50} and \ref{FigCoeffm2M50} respectively. The graphs illustrate that the method has an appropriate convergence rate.

\section{Conclusion}\label{Sec_Conclusion}
The Falkner–Skan non-linear ordinary differential equation arises in the study of laminar boundary layers. This problem is the subject of an extensive research to solve because of its importance in boundary layer theory, a solution for the non-linear ordinary differential laminar boundary layer Falkner-Skan equation has been obtained by pseudospectral method on semi-infinite domain, also in this work we concentrated on the boundary conditions of $f(0)=f'(0)=0,f'(+\infty)=1$, which is corresponding to a fixed and impermeable wedge flow.
The physical quantities of interest which is represented by the value of $f''(0)$ is the skin friction coefficient. According to the tables and figures value of $f''(0)$ can be seen to increase with $M$. Hence, the presence of a magnetic field also increases the skin friction.

Most numerical methods reported in the literature thus far are based on transforming maps of the physical domain $[0,+\infty)$ to the finite domains, shooting methods or finite-difference methods obtained by first truncating the semi-infinite physical domain of the problem to a finite domain at an unknown finite boundary, which is determined as part of the solution by imposing an ``asymptotic boundary condition" at this boundary.

The method presented in this paper used a set of Hermite functions and solved this problem on the semi-infinite domain without truncating it to a finite domain, imposing the asymptotic condition transforming and transforming domain of the problem. These functions are proposed to provide an effective but simple way to improve the convergence of the solution by pseudospectral method. The validity of the method is based on the assumption that it converges by increasing the number of Gauss points. Through the comparisons among the numerical solutions of Asaithambi and the current work, it has been shown that the present work has provided acceptable approach for Falkner-Skan equation; also it was confirmed by the theorem and logarithmic figures of absolute coefficients that this approach has exponentially convergence rate. In total, an important concern of spectral methods is the choice of basis functions; the basis functions have three number of properties: easy to computation, rapid convergence and completeness, which means that any solution can be represented to arbitrarily high accuracy by taking the truncation $N$ sufficiently large.



%% References

\section*{Acknowledgments}
The research was supported by a grant from Shahid Beheshti University.

\bibliographystyle{elsarticle-num}

\begin{thebibliography}{00}
\bibitem{Coulaud}O. Coulaud, D. Funaro, O. Kavian, %1---------------------
Laguerre spectral approximation of elliptic problems in exterior domains,
%Computer Methods in Applied Mechanics and Engineering.
Comput. Method. Appl. Mech. Eng.
80 1-3 (1990)451-458.

\bibitem{Funaro.Kavian} D. Funaro, O. Kavian, %2----------------------
Approximation of some diffusion evolution equations in unbounded domains by Hermite functions,
Math. Comp.
57 (1991)597-619.

\bibitem{Funaro.Appl. Numer. Math.1990} D. Funaro, %3-------------------
Computational aspects of pseudospectral Laguerre approximations,
%Applied Numerical Mathematics.
Appl. Numer. Math.
6 6 (1990)447-457.

\bibitem{Guo.Math. Comp.1999} B.Y. Guo, %4-----------------------------
Error estimation of Hermite spectral method for nonlinear partial differential equations,
%Mathematics of Computation.
Math. Comput.
68 227 (1999)1067-1078.

\bibitem{Guo.num2000}B.Y. Guo, J. Shen, %5---------------------------
Laguerre-Galerkin method for nonlinear partial differential equations on a semi-infinite interval,
%Numerische Mathematik.
Numer. Math.
86 4 (2000)635-654.

\bibitem{Maday} Y. Maday, B. Pernaud-Thomas, H. Vandeven, %6-------------------------
Reappraisal of Laguerre type spectral methods,
%La Recherche Aerospatiale.
La. Rech. Aerospatiale.
6 (1985)13-35.

\bibitem{Shen} J. Shen, %7--------------------------------
Stable and efficient spectral methods in unbounded domains using Laguerre functions,
%SIAM Journal on Numerical Analysis.
SIAM. J. Numer. Anal.
38 4 (2000)1113-1133.

\bibitem{Siyyam} H.I. Siyyam, %8--------------------------------
Laguerre Tau methods for solving higher order ordinary differential equations,
%Journal of Computational Analysis and Applications.
J. Comput. Anal. Appl.
3 2 (2001)173-182.

\bibitem{Guo.J. Math. Anal. Appl.1998}B.Y. Guo, %9----------------------------
Gegenbauer approximation and its applications to differential equations on the whole line,
%Journal of Mathematical Analysis and Applications.
J. Math. Anal. Appl.
226 1 (1998)180-206.

\bibitem{Guo.com2000} B.Y. Guo, %10-------------------------------
Jacobi spectral approximation and its applications to differential equations on the half line,
%Journal of Computational Mathematics.
J. Comput. Math.
18 1 (2000)95-112.

\bibitem{Guo.J. Math. Anal. Appl.2000}B.Y. Guo, %11------------------------------
Jacobi approximations in certain Hilbert spaces and their applications to singular differential equations,
%Journal of Mathematical Analysis and Applications.
J. Math. Anal. Appl.
243 2 (2000)373-408.
%doi:10.1006/jmaa.1999.6677

\bibitem{BoydBook} J.P. Boyd, %12-------------------------------
Chebyshev and Fourier Spectral Methods, second ed.,
Dover., New York, 2000.
%\book

\bibitem{Christov.SIAM J. Appl. Math.1982}CI. Christov, %13----------------------
A complete orthogonal system of functions in $L^2(-\infty,\infty)$ space,
%SIAM Journal on Applied Mathematics.
SIAM. J. Appl. Math.
42 (1982)1337-1344.

\bibitem{Boyd.J. Comput. Phys.1987(69)} J.P.Boyd, %14------------------------
Spectral methods using rational basis functions on an infinite interval,
%Journal of Computational Physics.
J. Comput. Phys.
69 1 (1987)112-142.

\bibitem{Boyd1987} J.P. Boyd, %15------------------------
Orthogonal rational functions on a semi-infinite interval,
%Journal of Computational Physics.
J. Comput. Phys.
70 1 (1987)63-88.

\bibitem{Guo.sci2000} B.Y. Guo, J. Shen, Z.Q. Wang, %16--------------------------
A rational approximation and its applications to differential equations on the half line,
%Journal of Scientific Computing.
J. Sci. Comput.
15 2 (2000)117-147.

\bibitem{Boyd2003} J.P. Boyd, C. Rangan, P.H. Bucksbaum, %17------------------------
Pseudospectral methods on a semi-infinite interval with application to the Hydrogen atom: a comparison of the mapped Fourier-sine method with Laguerre series and rational Chebyshev expansions,
%Journal of Computational Physics.
J. Comput. Phys.
188 1 (2003)56-74.
%doi:10.1016/S0021-9991(03)00127-X

\bibitem{Parand.Appl. Math. Comput.2004} K. Parand, M. Razzaghi, %18----------------------
Rational Chebyshev Tau method for solving Volterra's population model,
Appl. Math. Comput.
 149 3 (2004)893-900.

\bibitem{Parand.Int. J. Comput. Math.2004}K. Parand, M. Razzaghi, %19----------------------
Rational Chebyshev Tau method for solving higher-order ordinary differential equations,
%International Journal of Computer Mathematics.
Int. J. Comput. Math.
81 1 (2004)73-80.

\bibitem{Parand.Phys. Scripta2004} K. Parand, M. Razzaghi, %20-----------------------------
Rational Legendre approximation for solving some physical problems on semi-infinite intervals,
Phys. Scripta.
69 (2004)353-357.
%doi: 10.1238/Physica.Regular.069a00353

\bibitem{Parand.Phys.let.A} K. Parand, M. Shahini, %21-------------------------
Rational Chebyshev pseudospectral approach for solving Thomas-Fermi equation,
%Physics Letters A.
Phys. Lett. A.
373 (2009)210-213.

\bibitem{Parand.CAM} K. Parand, A. Taghavi, %22----------------------------
Rational scaled generalized Laguerre function collocation method for solving the Blasius equation,
%Journal of Computational and Applied Mathematics (2009),
J. Comput. Appl. Math.
233 4 (2009)980-989.
%doi:10.1016/j.cam.2009.08.106.

\bibitem{Parand.JCP} K. Parand, M. Shahini, M. Dehghan, %23----------------------------
Rational Legendre pseudospectral approach for solving nonlinear differential equations of Lane-Emden type,
%Journal of Computational Physics (2009),
J. Comput. Phys.
228 23 (2009)8830-8840.
%doi: 10.1016/j.jcp.2009.08.029

\bibitem{Schlichting}H. Schlichting,
Boundary Layer Theory, 6th ed., McGraw-Hill, New York, 1979. %------------------
%book

\bibitem{Ogulu.Makinde} A. Ogulu, O.D. Makinde,
Unsteady hydromagnetic free Convection flow of a dissipative and radiating fluid past a vertical plate with constant heat flux,
Chem. Eng. Comm.
196 (2009) 454–462.

\bibitem{MakindeHFF2009} O.D. Makinde,
On MHD boundary-layer flow and mass transfer past a vertical plate in a porous medium with constant heat flux,
Int. J. Numer. Method Heat Fluid Flow.
19 3/4 (2009) 546-554.

\bibitem{Makinde.Sibanda2008}O.D. Makinde, P. Sibanda,
Magnetohydrodynamic Mixed-Convective Flow and Heat and Mass Transfer Past a Vertical Plate in a Porous Medium With Constant Wall Suction,
J. Heat Transfer.
130 (2008) 1-8.

\bibitem{Falkner-Skan}V.M. Falkner, S.W. Skan,
Some approximate solutions of the boundary-layer equations, %----------
Phiols. Mag.
12 (1931)865-896.

\bibitem{Sutton-Sherman}G.W. Sutton, A. Sherman,
Engineering Magnetohydrodynamics, McGraw–Hill, %-------------------
New York, 1965.
%book

\bibitem{chiam}T.C. Chiam,
Hydromagnetic flow over a surface stretching with a power-law velocity, %------------------
Int. J. Eng. Sci.
33 (1995)429-435.

\bibitem{Yih}K.A. Yih, %-------------------------
MHD forced convection flow adjacent to a non-isothermal wedge,
Int. Comm. Heat. Mass. Transfer.
26 6 (1999)819-827.

\bibitem{ishak-nazar-pop}A. Ishak, R. Nazar, I. Pop,
MHD boundary-layer flow of a micropolar fluid past a wedge with constant wall heat flux,
Commun. Nonlinear. Sci. Numer. Simul.
14 (2009)109-118.

\bibitem{hayat-hussain-javad}T. Hayat, Q. Hussain, T. Javed,  %-----------------------
The modified decomposition method and Pad\'{e} approximants for the MHD flow over a non-linear stretching sheet,
Nonlinear. Anal. Real. World. Appl.
10 (2009)966-973.

\bibitem{Rashidi}M.M. Rashidi, %--------------------------
The modified differential transform method for solving MHD boundary-layer equations,
Comput. Phys. Commun.
180, (2009)2210-2217.

\bibitem{Addasbandi-Hayat1}S. Abbasbandy, T. Hayat, %------------------------
Solution of the MHD Falkner-Skan flow by homotopy analysis method,
Commun. Nonlinear. Sci. Numer. Simulat.
14 (2009)3591-3598.

\bibitem{Addasbandi-Hayat2}S. Abbasbandy, T. Hayat,
Solution of the MHD Falkner–Skan flow by Hankel–Pad\'{e} method,
Phys. Lett. A.
373 (2009)731-734.

\bibitem{GuoShenXu2003} B.y. Guo, J. Shen, J., C.l. Xu, %43-------------------------
Spectral and pseudospectral approximations using Hermite functions: application to the Dirac equation,
%Advances in Computational Mathematics.
Adv. Comput. Math.
19 1-3 (2003)35-55.

\bibitem{Bao.Shen}W. Bao, J. Shen, %44-------------------------------
A generalized-Laguerre-Hermite pseudospectral method for computing symmetric and central vortex states in Bose-Einstein condensates,
%Journal of Computational Physics.
J. Comput. Phys.
227 23 (2008)9778-9793.

\bibitem{Guo.Xu} B.Y. Guo, C.L. Xu, %45-------------------------------
Hermite pseudospectral method for nonlinear partial differential equations,
%Mathematical Modelling and Numerical Analysis.
Math. Model. Numer. Anal.
34 4 (2000)859-872.

\bibitem{ShenWang2008} J. Shen, L.L. Wang, %46---------------------------------
Some Recent Advances on Spectral Methods for Unbounded Domains,
%Communications in computational physics.
Commun. Comput. Phys.
5 2-4 (2009)195-241.

\bibitem{SzegoBook}G. Szeg$\ddot{o}$.
Orthogonal Polynomials, 4th edition,
volume 23. AMS Coll. Publ, 1975.

\bibitem{ShenTangHighOrder}J. Shen, T. Tang, %47------------------------------
High Order Numerical Methods and Algorithms,
Chinese Science Press, 2005.
%book

\bibitem{ShenTangWang}J. Shen, T. Tang, L-L. Wang,  %48
Spectral Methods Algorithms, Analyses and Applications,
Springer, First edition, 2010.
%book

\bibitem{Liu.Liu.Tang} Y. Liu, L. Liu, T. Tang, %49--------------------------
The numerical computation of connecting orbits in dynamical systems: a rational spectral approach,
J. Comput. Phys.
111 2 (1994)373-380.

\bibitem{Tang.1993}T. Tang, %50---------------------------
The Hermite spectral method for Gaussian-type functions,
SIAM. J. Sci. Comput.
14 3 (1993)594-606.

\bibitem{Asaithambi}N.S. Asaithambi, %---------------------------
A numerical method for the solution of the Falkner-Skan equation,
Appl. Math. Comput.
81 2-3 (1997)259-264.


\end{thebibliography}



%----------------------------------------------------------------------------------------------------------
\clearpage
\begin{table}
\caption{The comparison of HFP approximate solution of $f''(0)$ with numerical results \cite{Asaithambi} when m=-3/5}
\begin{tabular*}{\columnwidth}{@{\extracolsep{\fill}}*{6}{c}}
\hline
$M$ & $N$ & $k$ & $l$ & Present method & Numerical\\
\hline
$5$ & $20$ & $2$ & $1.658$ & $4.60075494$ & $4.60075494$\\
$10$ & $15$ & $2$ & $1.296$ & $9.80646420$ & $9.80646420$\\
$15$ & $15$ & $1$ & $1.089$ & $14.87167484$ & $14.87167484$\\
$20$ & $15$ & $1$ & $1$ & $19.90393701$ & $19.90393701$\\
$50$ & $20$ & $2$ & $1.336$ & $49.96165233$ & $49.96165233$\\
\hline
\end{tabular*}\label{Tab.Falkner-m3-5}
\end{table}
%----------------------------------------------------------------
\begin{table}
\caption{The comparison of HFP approximate solution of $f''(0)$ with numerical results \cite{Asaithambi} when $m=2$}
\begin{tabular*}{\columnwidth}{@{\extracolsep{\fill}}*{6}{c}}
\hline
$M$ & $N$ & $k$ & $l$ & Present method & Numerical\\
\hline
$5$ & $30$ & $3$ & $1.194$ & $5.19095945$ & $5.19095945$\\
$10$ & $30$ & $2$ & $1.112$ & $10.09677545$ & $10.09677545$\\
$50$ & $30$ & $2$ & $0.904$ & $50.01944071$ & $50.01944071$\\
$100$ & $30$ & $2$ & $0.616$ & $100.00972170$ & $100.00972170$\\
\hline
\end{tabular*}\label{Tab.Falkner-m2}
\end{table}
\clearpage

\clearpage

\end{document}